\documentclass{article}
\usepackage{INTERSPEECH2022}
\usepackage{xcolor}
\usepackage{amssymb}
\usepackage{amsmath,graphicx}
\usepackage{pdfpages}
\usepackage{hyperref}
\usepackage{booktabs}
\usepackage{multirow}
\usepackage{arydshln}
\usepackage{pifont}
\usepackage{hyperref}
\usepackage[hyphenbreaks]{breakurl}
\usepackage{subcaption}
\usepackage{caption}
\definecolor{aliceblue}{rgb}{0.94, 0.97, 1.0}
\definecolor{mistyrose}{rgb}{1.0, 0.49, 0.88}
 
\def\prepara{{\vspace{5pt}}}

\title{VoxSRC 2022: The Fourth VoxCeleb Speaker Recognition Challenge}

\name{\begin{tabular}{c}
Jaesung Huh$^{1}$, Andrew Brown$^1$, Jee-weon Jung$^{2}$, Joon Son Chung$^{1,3}$, Arsha Nagrani$^{1\dagger}$\footnotemark[2], \\ Daniel Garcia-Romero$^{4}$, Andrew Zisserman$^{1}$
\end{tabular}}

\address{$^1$Visual Geometry Group, Department of Engineering Science, University of Oxford, UK\\
$^2$Naver Corporation, South Korea \\
$^3$Korea Advanced Institute of Science and Technology, South Korea\\
$^4$AWS AI Labs, USA 
}
\email{\url{https://mm.kaist.ac.kr/datasets/voxceleb/voxsrc/competition2022.html}}

\begin{document}
\ninept
\maketitle
\begin{abstract}
This paper summarises the findings from the VoxCeleb Speaker Recognition Challenge 2022 (VoxSRC-22), which was held in conjunction with INTERSPEECH 2022. The goal of this challenge was to evaluate how well state-of-the-art speaker recognition systems can diarise and recognise speakers from speech obtained ``in the wild''. The challenge consisted of: (i) the provision of publicly available speaker recognition and diarisation data from YouTube videos together with ground truth annotation and standardised evaluation software;  and (ii) a public challenge and hybrid workshop held at INTERSPEECH 2022. We describe the four tracks of our challenge along with the baselines, methods, and results. We conclude with a discussion on the new domain-transfer focus of VoxSRC-22, and on the progression of the challenge from the previous three editions.
\end{abstract}

\renewcommand*{\thefootnote}{\fnsymbol{footnote}}
\footnotetext[2]{Also at Google Research.}


\section{Introduction}
\label{sec:intro}
The fourth edition of the VoxCeleb Speaker Recognition Challenge was held in 2022 (VoxSRC-22). The main objectives of this series are to: (i) investigate and advance new speaker recognition research ``in the wild''; (ii) gauge and calibrate the performance of current technology through open evaluation tools; and (iii) provide open-source data that is available to all members of the research community.

Each year, VoxSRC introduces a new special focus. In the second installation (VoxSRC-20~\cite{nagrani2020voxsrc}), we introduced two new tracks: (i) the self-supervised verification track (inspired by the successes in self-supervised learning~\cite{nagrani2020disentangled, huh2020augmentation}), where no speaker labels can be used during the pretrain phase; and (ii) a speaker diarisation track which exploits the VoxConverse~\cite{Chung20} dataset. In the third edition (VoxSRC-21~\cite{brown2022voxsrc}), we added a multi-lingual focus to the verification tracks, to encourage fairness and diversity and to build a more challenging test set.

This year, we introduced a new track focused on semi-supervised domain adaptation. The goal was to assess how models pretrained on large labelled data in a {\em source} domain can adapt to a new \textit{target} domain, given (i) a large set of unlabelled data from the target domain and (ii) a small set of labelled data from the target domain.
This is especially relevant and important to low resource real-world scenarios, where large scale labelled data is not available in a target domain, but a sufficiently large dataset from another domain such as VoxCeleb~\cite{Nagrani17} is available. 

For the existing speaker verification tracks, we also applied two novel techniques for making more challenging positive and negative pairs for the speaker verification test set, by using a face age classifier and a speaker diarisation dataset, respectively.

This paper details the four evaluation tasks, provided datasets, the submissions, and winners of VoxSRC-22 challenge. Please refer to our website for more information. 

\section{Task Description} 
\subsection{Tracks} \label{sec:tracks}
The challenge consisted of the following four tracks:
\begin{enumerate}
    \item Speaker Verification (Closed)
    \item Speaker Verification (Open)
    \item Semi-supervised domain adaptation (Closed)
    \item Speaker diarisation (Open)
\end{enumerate}
For the verification tracks, the open and closed training conditions refer to the training data that is allowed. 
The tasks of Tracks 1, 2, and 4 were identical to those of last year's challenge, whereas the track 3 task of semi-supervised domain adaptation was newly introduced this year. 
Please see the following section for further details.

\subsection{Data}
\label{subsec:data}
\subsubsection{Speaker Verification -- Track 1 and 2}
\label{subsubsec:data_track12}
The VoxCeleb datasets~\cite{Nagrani17, nagrani2020voxceleb,Chung18a} contain speech utterances from YouTube videos, including celebrity interviews and TV shows. 
Please refer to~\cite{nagrani2020voxceleb} for more detailed descriptions.


\prepara\noindent\textbf{Training sets:} 
For track 1 (closed) participants were permitted only to use the VoxCeleb2 dev set~\cite{Chung18a}, which contains more than a million utterances from 5,994 speakers. 
For track 2 (open), participants were permitted to use any other external datasets in addition to the VoxCeleb2 dev set for training, but not the challenge's test data.

\prepara\noindent\textbf{Validation and Test sets:} 
This year, we focused on making the validation and the test sets more challenging by introducing two new trial types -- hard positives and hard negatives.

We constructed hard positives where the age of the speaker differs considerably between the two utterances. The hard positives were found by selecting utterance pairs from the same speaker that have a large age gap (i.e.\ two audio files for the same identity where the age is very different) via a two step process on private VoxCeleb video data. In this VoxCeleb data, for each video segment we have the face, identity and speech.
First, the age of the speaker is estimated by predicting the age for a random set of frames, using
an open-source age prediction network~\cite{FaceLibGit}, and averaging the result. 
Second, we sample positive pairs from utterances for the same speaker with large age gaps.

We constructed hard negatives using utterances from the same video.
When sampling a negative pair using utterances from different videos, speaker verification systems may be able to rely on cues from the different microphones or room environments to help discriminate the different identities of the speakers, which can make the task easier. 
Our goal here was therefore to construct harder negative pairs by sampling utterances from different speakers that are from the same audio file. 
In this case, the microphone and environment noise are shared across the two utterances, and only the identity of the speaker changes. 
We sampled the hard negative pairs using speaker diarisation datasets, where each audio file consists of multiple short speech segments from different speakers.
To generate these trials, we first cropped short speech segments.
We then removed segments that are either too short ($<$1.5s) or contains overlapping speech.
Finally, we selected trials using two segments within an audio file.
Full details are given in~\cite{jung2022search}. 

We released 305,196 validation pairs and 317,973 test pairs, including these hard positives and negatives. 
We also included the VoxSRC-19 test pairs in our test set to track the state-of-the-art performance on the same trials. 
No overlapping speakers exist between validation set and test set. 
Statistics of the val / test sets are reported in Table~\ref{tab:testdata_track123}.

\subsubsection{Semi-supervised domain adaptation -- Track 3}
\label{subsubsec:data_track3}
This year, we introduced a new track focused on semi-supervised domain adaptation. 
Here, we focused on the problem of how models, pretrained on a large set of data with labels in a source domain, can adapt to a new target domain given: (i) a large set of unlabelled data from the target domain, and (ii) a small set of labelled data from the target domain.
Specifically, the domain adaptation that we focused on is from one language in a source domain (mainly English), to a different language in a target domain (Chinese), for the task of speaker verification. 
Here we use VoxCeleb~\cite{nagrani2020voxceleb} for the source domain, and CN-Celeb~\cite{li2022cn} for the target domain. 

\prepara\noindent\textbf{Train set:}
Participants were allowed to use three types of datasets in this track:

\begin{itemize}
  \item VoxCeleb2 dev set \textit{with} speaker labels (Source domain). This can be used for pretraining.
  \item A large subset of CN-Celeb \textit{without} speaker labels (Target domain). This can be used for domain adaptation.
  \item A small subset of CN-Celeb \textit{with} speaker labels (Target domain) consisting of 20 utterances each from 50 different speakers.
\end{itemize}


VoxCeleb2 data consists mainly of interview-style utterances, whereas CN-Celeb consists of several different genres. 
To focus on the language domain adaptation task, we have therefore removed utterances in the ``singing'', ``play'', ``movie'', ``advertisement'', and ``drama'' genres from CN-Celeb.

\prepara\noindent\textbf{Validation and Test sets:} 
For the validation and test set, we provided a list of trial speech pairs from identities in the target domain. 
We created and released a validation set consisting of 40,000 validation pairs. 
The test set consists of 30,000 pairs from disjoint identities not present in either CN-Celeb1 or CN-Celeb2. 
Each trial contains two single-speaker speech segments, of variable length. 
See Table~\ref{tab:testdata_track123} for detailed statistics.

\subsubsection{Speaker Diarisation -- Track 4}
\label{subsubsec:data_track4}
VoxConverse~\cite{Chung20} is a speaker diarisation dataset from diverse domains such as panel discussions, news segments and talk shows. 
It consists of multi-speaker audio segments with challenging background conditions and overlapping speech. 
Please refer to \cite{Chung20} for more details.

\prepara\noindent\textbf{Training set:} 
Similar to previous years, participants were allowed to train their models on \textit{any} data, except for the test set of the challenge.

\prepara\noindent\textbf{Validation set:} 
Participants were allowed to use both dev / test sets of the VoxConverse dataset. 
The total duration of VoxConverse is approximately 64 hours, and the average number of speakers per audio segment ranges between 4 and 6. 
The average percentage of speech per each audio file is 91\%.

\prepara\noindent\textbf{Test set:} 
The test set contains 360 audio files, created with the identical semi-automatic pipeline used for creating VoxConverse. 
The Track 4 VoxSRC-2021 test set is included as a subset of the test set. 
In addition, we included an additional 96 audio files from YouTube videos in diverse categories, including news, documentary, lecture and commercial. 
Details for both validation and test set are described in Table~\ref{tab:testdata_track4}.



\begin{table}[]
    \centering
    \begin{tabular}{cccccc}
    \toprule
         \textbf{Track}& \textbf{Split} & \textbf{\# Pairs} & \textbf{\# Utter.} & \textbf{Segment length (s)}  \\ \midrule
         \multirow{2}{*}{1 \& 2} & val  & 305,196 & 110,366 & 2.00 / 8.43 / 314.44\\
         & test & 317,973 & 34,684 & 1.98 / 7.36 / 282.16\\
         \midrule
          \multirow{2}{*}{3} & val  & 40,000 & 2,400 & 0.44 / 9.38 / 224.65\\
         & test & 30,000 & 18,377 & 1.23 / 9.06 / 89.83\\
         \bottomrule
    \end{tabular}
    \caption{\small{Statistics of the speaker verification validation and test sets (Tracks 1--3). \textbf{\# Pairs} refers to the number of evaluation trial pairs, whereas \textbf{\# Utter.} refers to the total number of unique speech segments. Segment lengths are reported as min/mean/max.}}
    \label{tab:testdata_track123}
\end{table}

\begin{table}[]
    \centering
    \footnotesize
    \resizebox{1\linewidth}{!}{
    \begin{tabular}{ccccc}
    \toprule
         \textbf{Split} & \textbf{\# audios} & \textbf{\# spks} & \textbf{Duration (s)} & \textbf{speech \%} \\ \midrule
         val  & 448 & 1 / 5.5 / 21 & 22.0 / 512.9 / 1200.0 & 11 / 91 / 100 \\
         test & 360 & 1 / 5.5 / 28 & 27.5 / 449.2 / 1777.8 & 9 / 88 / 100 \\
         \bottomrule
    \end{tabular}}
    \caption{\small{Statistics of the speaker diarisation val and test sets (Track 4). 
    Entries that have 3 values are reported as min/mean/max. \textbf{\#~spks:} Number of speakers per video. \textbf{Duration (s):} Length of videos in seconds. \textbf{speech \%:} Percentage of video time that is speech.}}
    \label{tab:testdata_track4}
    \normalsize
\end{table}

\begin{table*}[]
    \renewcommand\arraystretch{1.1}
    \centering
    
    \begin{tabular}{cccccc}
    \toprule
         \textbf{Track} & \textbf{Rank} & \textbf{Team Name} & \textbf{Organisation} & \textbf{minDCF} & \textbf{EER}  \\ 
          \midrule
       \multirow{3}{*}{1}  & 3 & SJTU-AISPEECH~\cite{chen2022sjtu} & Shanghai Jiao Tong University, AISpeech Ltd & 0.101 & 1.911 \\ 
         & 2 & KristonAI~\cite{cai2022kriston} & KristonAI Lab  & 0.090 & 1.401\\ 
        & 1 & ravana - ID R\&D ~\cite{ravana2022} & ID R\&D Lab & 0.088 & 1.486\\ 
         \midrule
        \multirow{3}{*}{2} & 3 & Strasbourg-Spk & Microsoft & 0.073 & 1.436 \\ 
         & 2 & KristonAI~\cite{cai2022kriston} & KristonAI Lab  & 0.072 & 1.119 \\ 
        & 1 & ravana - ID R\&D ~\cite{ravana2022} & ID R\&D Lab & 0.062 & 1.212\\ 
         \cmidrule{1-6}\morecmidrules\cmidrule{1-6}
         \multirow{3}{*}{3}  & 3 & SJTU-AISPEECH~\cite{chen2022sjtu} & Shanghai Jiao Tong University, AISpeech Ltd & 0.437 & 8.087 \\ 
         &2 & DKU-Tencent~\cite{qin2022dku} & Duke Kunshan University, Tencent AI Lab & 0.389 & 7.153 \\ 
  &1 & zzdddz~\cite{zhaohccl2022} & Chinese Academy of Sciences & 0.388 & 7.030\\ 
         \bottomrule
    \end{tabular}
    
    \caption{\small{Winners for the speaker verification tracks (Tracks 1, 2 and 3). The primary metric for Track 1 \& 2 is \textbf{minDCF} while the primary metric for Track 3 is \textbf{EER}. For both metrics, a lower score is better. Note that Track 1 \& 2 have an identical test set.}}
    \label{tab:results_verification}
\end{table*}

\begin{table*}[]
    \renewcommand\arraystretch{1.1}
    \centering
    \begin{tabular}{ccccc}
    \toprule
         \textbf{Rank} & \textbf{Team Name} & \textbf{Organisation} & \textbf{DER} & \textbf{JER}  \\ \midrule
         3 & AiTER~\cite{park2022gist} & Gwangju Institute of Science and Technology & 5.12 & 30.82\\ 
         2 & KristonAI~\cite{cai2022kriston} & KristonAI Lab  & 4.87 & 25.49 \\ 
         1 & DKU-DukeECE~\cite{wang2022dku} & Duke Kunshan University, Duke University & 4.75 & 27.85\\ 
         \bottomrule
    \end{tabular}
    \caption{\small{Winners for the speaker diarisation track (Track 4). The primary metric is \textbf{DER}. For both metrics, a lower score is better.}}
    \label{tab:results_diarsiation}
\end{table*}

\section{Challenge Mechanics} 

\subsection{Evaluation metrics}
A validation toolkit\footnote{\url{https://github.com/JaesungHuh/VoxSRC2022}} was provided for both speaker verification and speaker diarisation. 
Participants were advised to test their models on the validation set for each track using this open-sourced code. 
The evaluation metrics are identical to VoxSRC 2021~\cite{brown2022voxsrc}.

\prepara\noindent\textbf{Speaker verification.}
We reported two evaluation metrics: (i) the Equal Error Rate (EER) which is a location on a ROC or DET curve where the false acceptance rate and false rejection rate are equal; and (ii) minDCF ($C_{DET}$) used by the previous VoxSRC~\cite{brown2022voxsrc, chung2019voxsrc} evaluations. 
We used $C_{miss} = C_{fa} = 1$ and $P_{tar}=0.05$ in our cost function. 
The main metric for Tracks 1 and 2 was minDCF, and the final ranking was based only on this score. 
EER was used as the main metric for Track 3.

\prepara\noindent\textbf{Speaker diarisation.}
We chose two diarisation metrics, Diarisation Error Rate (DER) and Jaccard Error Rate (JER).
DER is the sum of speaker error, false alarm speech and missed speech. 
We used a 0.25-second forgiving collar, and overlapping speech was not disregarded. 
JER is based on the Jaccard index, which is defined as the ratio between the intersection and union of two segmentations. 
It is computed as a 1 minus the average of Jaccard index of optimal mappings between reference and system speakers~\cite{ryant2019second}.

\subsection{Baselines} \label{baselines}
\label{subsec:baselines}
We used same the baseline models as for last year's challenge, so please refer to~\cite{brown2022voxsrc} for more details.

We used the publicly released speaker verification network trained only with VoxCeleb2 dev set for verification tracks~\cite{kwon2021ins}. 
The model is ResNet-34~\cite{he2015deep} with ASP pooling~\cite{okabe2018attentive} and is trained with a combination of angular prototypical loss~\cite{chung2020defence} and cross-entropy loss.
This baseline achieved a minDCF of 0.346 and an EER of 5.63\% on track 1 and 2 test pairs, but a minDCF of 0.823 and an EER of 16.9\% on track 3 test pairs. 
This performance gap shows the necessity of domain adaptation on different language utterances.

For the diarisation track, we adopted a system described in~\cite{wang2018speaker} using a speaker embedding extractor to our baseline speaker model, publicly available py-webrtcvad~\cite{pywebrtc}, and an agglomerative hierarchical clustering (AHC) of speaker representation. 
The resulting model achieved 19.6\% DER and 41.4\% JER on the challenge test set.

\subsection{Submission} 
The challenge was hosted based on publicly available CodaLab code \footnote{\url{https://github.com/codalab/codalab-competitions}}, but hosting on our own evaluation instance for efficient maintenance. 
Similar to last year, we introduced two phases: ``Challenge workshop'' and ``Permanent'' and the challenge results were based on the former phase. 
Participants could only submit one submission per day and ten submissions in total. 
Submission for the ``Challenge workshop" phase was available until 14$^{th}$ of September, 2022.
Participants were required to submit reports of their methods and results by 20$^{th}$  of September 2022. 

\section{Workshop}
\label{sec:workshop}
VoxSRC-22 was a hybrid workshop with both in-person and virtual attendance options. 
The in-person workshop was held on the 22$^{nd}$ of September in Incheon Songdo Convensia, the conference venue of INTERSPEECH 2022. 
The workshop was free of cost for attendees. 

The workshop began with an introductory talk from the organisers, followed by a keynote speech from professor Junichi Yamagishi, titled ``The use of speaker embeddings in neural audio generations''. 
The winners then gave short presentations about their methods and results. 
All slides and presentation videos are available on our workshop website\footnote{\url{https://mm.kaist.ac.kr/datasets/voxceleb/voxsrc/interspeech2022.html}}. 

\section{Methods and Results} 
\label{method_and_results}

There were a total of 554 submissions across all four tracks this year. 
The performances of the top three ranked teams for each track are reported in Table~\ref{tab:results_verification} and Table~\ref{tab:results_diarsiation}, along with their scores.
In this section, we give details on the methods used by the top two ranked teams from each track.

\subsection{Speaker Verification (Track 1 and 2)}
\label{subsec:method_track12}
This year, Tracks 1 and 2 had the same winners and runner-up.
The winning team~\cite{ravana2022} adopted a fusion of deep ResNets and ECAPA-TDNN~\cite{desplanques2020ecapa} along with extensive data augmentations using the MUSAN noise database~\cite{snyder2015musan} and RIR~\cite{ko2017study} responses. 
For Track 2, they trained the model with their own \textit{Self-VoxCeleb} dataset inspired by the data collection pipeline from VoxCeleb but only using speech-based filtering. 
The inclusion of Self-VoxCeleb improved 20-50\% of relative performance compared to the models trained only with VoxCeleb2. 
AS-Norm and QMF functions were employed for post-processing the scores. 

The second-place~\cite{cai2022kriston} employed ResNet variants with diverse input features, model depths and kernel sizes in Track 1. 
Data augmentation was carried out using the MUSAN noise database and RIR responses, which are similar to the winner's method.
Moreover, they applied 3-fold speed augmentation to enlarge the training dataset, resulting in obtaining 17,982 speakers.
For Track 2, they utilised several recently proposed pretrained networks, such as WavLM~\cite{chen2022wavlm} and variants of Wav2Vec2~\cite{baevski2020wav2vec} and ensembled these networks with the models that they trained for Track 1.
All of their models followed two-step training, training the model only with short utterances followed by training the model including longer ones with large margin fine-tuning.
Their submission performed better than the winning team in terms of EER, but performed slightly worse in minDCF, which is our primary metric.

\prepara\noindent\textbf{Effect of self-supervised speaker models.}
This year, the top two winning teams in Track 2 obtained impressive performance gains by utilising models trained with self-supervision on large scale data, which had not been observed in previous additions. 

Following the great success in the fields of vision~\cite{chen2020simple, grill2020bootstrap, zbontar2021barlow} and NLP~\cite{devlin2018bert, yang2019xlnet, lan2019albert}, self-supervised learning has also shown prominent results in speech processing~\cite{chen2022wavlm,baevski2020wav2vec, hsu2021hubert}.
Since the supervised training of speech networks with labels and annotations disregards rich information in the input signal, self-supervised methods instead enable the model to learn a universal representation, such as speaker information. 
The winner~\cite{ravana2022} utilised WavLM~\cite{chen2022wavlm} and Hubert~\cite{hsu2021hubert} pretrained models and achieved 30\% relative improvement on minDCF, our primary metric.
The second place leveraged pretrained WavLM~\cite{chen2022wavlm} and Wav2Vec2~\cite{baevski2020wav2vec} and finetuned them with VoxCeleb before fusing with other models.
They also achieved 20\% relative improvement on our primary metric (0.090 to 0.062).

\prepara\noindent\textbf{Analysis on hard positive and negative pairs.}
This year we introduced new trial types to make the test set harder, as described in Section~\ref{subsubsec:data_track12}.
Here we analyse how these pairs affect the winners' performance.
The VoxSRC-22 test set consists of four types of trials, (i) hard positive pairs taken from the same speaker at different ages (\textbf{P-H}), (ii) hard negative pairs taken from the same environment (\textbf{N-H}), (iii) positive pairs from VoxSRC-19 test set (\textbf{P-Vox19}), and (iv) negative pairs from VoxSRC-19 test set (\textbf{N-Vox19}).
We compare the performance of our baseline model and the top 2 winners of track 1 on these subsets.

Table~\ref{tab:track12_analysis} shows the results. 
The 1st~\cite{ravana2022} and 2nd place~\cite{cai2022kriston} performed better than our baseline model by a large margin. 
Comparing the performance of E-1 to the others shows that both the hard positives and the hard negatives made the challenge more difficult.
For the most challenging set, E-4 with both hard positive and negative pairs, the 1st place method (which achieves an impressive 0.9 \% EER on the VoxSRC-19 test set) could only achieve 2.07\% on the E-4 eval set.
Interestingly, the 2nd place method performed better in E-1, E-2 and E-3 than the 1st place but achieved worse results in E-4.
In fact, there is not much difference in overall performance between the first and second placed methods (See Table~\ref{tab:results_verification}).

\begin{table}[t!]
    \centering
    \begin{tabular}{cccccc}
    \toprule
\begin{tabular}[c]{@{}c@{}}\textbf{Eval.} \\ \textbf{set}\end{tabular} & \begin{tabular}[c]{@{}c@{}}\textbf{Positive} \\ \textbf{Pairs}\end{tabular} & \begin{tabular}[c]{@{}c@{}}\textbf{Negative} \\ \textbf{Pairs}\end{tabular} & \begin{tabular}[c]{@{}c@{}}\textbf{Baseline}\end{tabular}  &  \begin{tabular}[c]{@{}c@{}}\textbf{1st} \\ \textbf{place}\end{tabular}  &  
\begin{tabular}[c]{@{}c@{}}\textbf{2nd} \\ \textbf{place}\end{tabular}  \\\midrule
        E-1 & \textbf{P-Vox19} & \textbf{N-Vox19} & 1.47  &  0.90  & 0.65 \\
        E-2 & \textbf{P-Vox19} & \textbf{N-H} & 3.25  &  1.35  & 1.15 \\
        E-3 & \textbf{P-H} & \textbf{N-Vox19} & 4.50  &  1.33  & 1.18 \\
        E-4 & \textbf{P-H} & \textbf{N-H} & 9.27  &  2.07  & 2.28 \\
    \bottomrule
    \end{tabular}
    \caption{\small{Performance of baseline model and winning methods in Track 1 on four subsets of the test set. We report \% EER. Lower is better. \textbf{P-Vox19} : positive pairs from VoxSRC-19 test set, \textbf{N-Vox19} : negative pairs from VoxSRC-19 test set, \textbf{P-H} : hard positive pairs taken from same speaker at different ages, and \textbf{N-H} : hard negative pairs taken from the same environment. }}
    \label{tab:track12_analysis}
\end{table} 

 \begin{table}[]
    \centering
    \begin{tabular}{cccc}
    \toprule
         \textbf{Model} & \textbf{Train dataset} & \textbf{minDCF} &\textbf{EER}  \\ \midrule
        Baseline 1 & \textbf{L-S} & 0.823 & 16.88 \\
        Baseline 2 & \textbf{L-T} & 0.999 & 32.47 \\
        Baseline 3 & \textbf{L-S} + \textbf{L-T} & 0.687 & 13.93\\
        \midrule
        1st place~\cite{zhaohccl2022} & \textbf{L-S} + \textbf{U-T} + \textbf{L-T} & 0.388 & 7.03 \\
        2nd place~\cite{qin2022dku} & \textbf{L-S} + \textbf{U-T} + \textbf{L-T} & 0.389 & 7.15 \\
    \bottomrule
    \end{tabular}
    \caption{\small{Comparison of winning methods in Track3 with baselines. \textbf{L-S} : Labelled data in Source domain, \textbf{U-T}: Unlabelled data in Target domain and \textbf{L-T} Labelled data in Target domain.}}
    \label{tab:track3_analysis}
\end{table} 

\subsection{Semi-Supervised Domain Adaptation (Track 3)}
\label{subsec:method_track3}

The first placed team~\cite{zhaohccl2022} used two frameworks, pseudo labelling and self-supervised learning, to achieve the winning performance on the target domain. 
A novel sub-graph clustering algorithm based on two Gaussian fitting and multi-model voting was used for generating pseudo-labels. 
The model was trained with two stages, first using the labelled source domain data and pseudo-labelled target domain data, and second finetuning CN-Celeb data by fixing the VoxCeleb weights of the classification layer using circle loss. 
Then the pseudo label correction method was adopted and the model was retrained with them. 
They also tried various types of domain adaptation techniques, such as CORAL~\cite{sun2017correlation} or CORAL+~\cite{lee2019coral+}, but the performance did not improve.

The second placed team~\cite{qin2022dku} followed the FFSVC baseline system method~\cite{qin2020interspeech}. 
The clustering-based method wass used to generate the pseudo-labels of unlabelled target domain data. 
They used the track 1 speaker model as their initial checkpoint and finetuned with CN-Celeb data and pseudo labels.
Sub-center ArcFace was used for the loss function which was persistent with noisy labels. 
QMF-based score calibration and score normalisation were used as post-processing steps.

\prepara\noindent\textbf{Discussion.}
Table~\ref{tab:track3_analysis} shows the  top two teams' performance on the test set compared to several baselines. 
We trained three baseline models using same architecture and loss functions described in Section~\ref{subsec:baselines} but with different training sets. 
\textbf{Baseline 1} is identical to the baseline model for Tracks 1 and 2 which was trained only with the VoxCeleb2 dev set, the labelled data in the \textit{source} domain (\textbf{L-S}).
We also provide \textbf{Baseline 2}, which was trained only with the labelled data in the \textit{target} domain (\textbf{L-T}) from scratch. 
\textbf{Baseline 3} was trained starting from Baseline 1 and finetuned with labelled data in the target domain using a low learning rate (1e-5).
None of these baselines utilised the large amounts of unlabelled target domain data that was available to participants.

A comparison on Baseline 1 and 3 shows that including the labelled data in the target domain results in a performance improvement, relatively 2\% in terms of EER, even though the size of the labelled target domain data is negligible. 
However, Baseline 2 shows that using only the labelled target domain data results in a substantial performance decrease due to  over-fitting. 
Finally, the two winners' performances show that utilising the extensive unlabelled target domain data is essential for performance improvement in the train set, such as in the form of pseudo-labelling during training. 


\subsection{Speaker diarisation (Track 4)}
Track 4 saw 101 submissions from 17 different teams this year. 
The performances of the top three ranked teams are shown in Table~\ref{tab:results_diarsiation}.

The winner~\cite{wang2022dku} of this track employed a similar approach to their previous year's system~\cite{wang2021dku} in VoxSRC-21.
They adopted several conventional clustering-based diarisation system pipelines, which were fused using DOVER-LAP~\cite{raj2021dover}.
The differences from their last year's submission are two-fold.
First, they adopted a better speaker embedding extractor to bridge the domain gap between VoxCeleb and VoxConverse.
Second, they used four different voice activity detection models, ResNet-based, Conformer-based, VAD from \textit{pyannote.audio 2.0}~\cite{bredin2020pyannote} and ASR-based VAD, and performed majority voting from the results of these models. 
The winner achieved 4.57\% DER on the challenge test set.

The second place~\cite{cai2022kriston} team also adopted a conventional clustering-based system pipeline.
They re-trained the VAD models explained in~\cite{wang2021dku} but with different acoustic features, including 30-dim MFCC and 80-dim filterbank, and fused the results with \textit{pyannote.audio 2.0}.
A speaker embedding extractor, also used in their track 1 submission that achieves an EER 0.44\% in VoxCeleb1-O has been employed.
They applied two steps of clustering, initially with AHC, followed by a re-clustering step using a Bayesian hidden Markov model.
Unlike the winning team, they employed an additional module for handling overlapped speech where its training process was similar to that of their VAD models. 
They assigned the two most likely speakers to the overlapping speech.


 \begin{table}[]
    \renewcommand\arraystretch{1.1}
    \centering
    \resizebox{1\linewidth}{!}{
    \begin{tabular}{lccccc}
    \toprule
         \textbf{Team} & \textbf{Track} & \textbf{2019 test} & \textbf{2020 test}   & \textbf{2021 test} & \textbf{2022 test} \\ \midrule
         VoxSRC 2019 winner~\cite{zeinali2019but} & 1 & 1.42 & - & - & -\\
         \midrule
          VoxSRC 2020 winner ~\cite{thienpondt2020idlab} & 1,2 & 0.80 &  3.73 & - & -\\  
          VoxSRC 2020 2nd place~\cite{xiang2020xx205} & 1,2 & 0.75 & 3.81 & - & -\\
         \midrule
         VoxSRC 2021 winner~\cite{zhao2021speakin} & 1,2 & 0.57 & - & 1.85 & -\\  
          VoxSRC 2021 2nd place~\cite{zhang2021beijing} & 1,2 & 0.62 & - & 2.84 & -\\
        \midrule
         VoxSRC 2022 winner~\cite{ravana2022} & 1 & 0.90 & - & - & 1.49 \\  
          VoxSRC 2022 2nd place~\cite{cai2022kriston} & 1 & 0.65 & - & - & 1.40 \\
        \cmidrule{1-6}\morecmidrules\cmidrule{1-6}
          VoxSRC 2022 winner~\cite{ravana2022} & 2 & 0.69 & - & - & 1.21 \\  
          VoxSRC 2022 2nd place~\cite{cai2022kriston} & 2 & \textbf{0.50} & - & - & 1.12 \\
          \bottomrule
    \end{tabular}}
    \caption{\small{Comparison of methods (\% EER) on the four workshop test sets of Track 1 and 2. The 2019 test set is contained in the test sets of 2020, 2021 and 2022, meaning performance can be compared via the 2019 test set. We compare the VoxSRC 2019 winning submission and the top-2 submissions from both VoxSRC 2020 and VoxSRC 2021 on the 2019 test set, showing the large performance improvement in a year. For \% EER shown, lower is better.}}
    \label{tab:results_comparison_verification}
\end{table} 

\begin{table}[]
    \centering
    \footnotesize
    \resizebox{1\linewidth}{!}{
    \begin{tabular}{lccccc}
    \toprule
     & &\multicolumn{2}{c}{\textbf{2021 test}} & \multicolumn{2}{c}{\textbf{2022 test}} \\
     \midrule
    \textbf{Team} & \textbf{Track} & \textbf{DER}   & \textbf{JER}   & \textbf{DER}    & \textbf{JER}             \\ \midrule
        VoxSRC 2021 winner~\cite{wang2021dku}& 4 &      5.07           &         29.16      &           -      &    -    \\
         VoxSRC 2021 second place~\cite{wang2021bytedance}& 4 &5.15        &          26.02    &      -      &    -    \\
         \midrule
        VoxSRC 2022 winner~\cite{wang2022dku}&  4 &        4.16       &    24.75     &     4.75            &       27.85        \\
         VoxSRC 2022 second place~\cite{cai2022kriston}& 4 &  4.05     &   21.73      &        4.87         &    25.49            \\
     \bottomrule
    \end{tabular}}
    \caption{\small{Comparison of methods on the test sets of Track 4. We report the performance of the top two teams on the 2021 test set, demonstrating the performance improvement in a year. For both \% metrics, lower is better.}}
    \label{tab:results_comparison_diarisation}
    \normalsize
\end{table} 

\section{Discussion} \label{sec:discussion}

This year, the number of workshop participants was high because we offered two options for participation, in-person or virtual attendance.
50 participants attended in person, and 100 participants attended virtually on average.
For the winners' talks, the virtual attendees sent the organisers pre-recorded videos which explained their methods and results, while the in-person attendees gave their talks in the workshop venue.
Questions were collected from both Zoom and people who attended in-person.
All the slides and recorded talks are now available on our website.

For Track 3 which was newly introduced this year, we received a large number of submissions: 89 submissions from 42 participants.
This indicates a great interest from the speaker verification community in building methods for bridging the gap between two different domains.
The winning methods here leveraged pseudo-labelling using a model pretrained on the abundant labelled source domain data.
Interestingly, the winning methods did not see performance boosts from using classic domain adaptation techniques such as CORAL~\cite{sun2017correlation}.
Explanations for this could be the recent availability of powerful self-supervised speaker models, which have been shown to achieve outstanding speaker verification performance in recent years.
We hope more methods specific to domain adaptation could be explored next year.

We have included the entire VoxSRC-19 test set in the verification test set every year, allowing us to compare the techniques from previous rounds of the challenge.
Table~\ref{tab:results_comparison_verification} shows the improvement in the challenge-winning methods over the last four years.
By comparing the Track 1 winning methods over the years, we see that this year the winners' performance were slightly worse than in both the 2020 and 2021 editions.
However, when comparing Track 2 submissions, the performance improved significantly, possibly due to the inclusion of the self-supervised trained speaker models, such as WavLM~\cite{chen2022wavlm} or Wav2Vec~\cite{baevski2020wav2vec}.
Note that the 2nd place performs better on EER than the first place this year but performs worse on the Detection Cost Function (minDCF), which is our primary metric.

For the diarisation track, we also included the VoxSRC-21 test set in the VoxSRC-22 test set to assess the performance of all submissions made this year based solely on the 2021 test set. 
Table~\ref{tab:results_comparison_diarisation} shows the result.
Comparing the winners' performance on VoxSRC-21 test set, we see a significant performance improvement of state-of-the-art methods in a year. (DER 5.07\% vs 4.16\%).
We also compare the performance between top-2 winners' submissions on the 2021 and 2022 test sets.
It demonstrates that this year's test set is more challenging than last year's.
Somewhat surprisingly, the second place~\cite{cai2022kriston} achieved better performance on the 2021 test set compared to the winner.

\vskip 3mm

\noindent{\bf Acknowledgements.} We thank the authors of CN-Celeb for their help and support. 
We also thank Rajan from Elancer and his team, \url{http://elancerits.com/}, for their huge assistance with diarisation annotation, Kihyun Nam, Doyeop Kwak and Youngjoon Jang for double-checking the diarisation test set labels and David Pinto for supporting the evaluation server. 
We are grateful to Mitchell McLaren and Doug Reynolds for their continued support to VoxSRC. 
This work is funded by the EPSRC programme grant EP/T028572/1 VisualAI. 
Jaesung Huh is funded by a Global Korea Scholarship. 
We are grateful to Naver for sponsoring the workshop.


\bibliographystyle{IEEEtran}
\bibliography{shortstrings,refs}

\end{document}